\documentstyle[graphicx,aps]{revtex}

\begin{document}
\draft

 \twocolumn[\hsize\textwidth\columnwidth\hsize  
 \csname @twocolumnfalse\endcsname              


 



\title{
Comment on ''Experimental observation of the topological structure of
exceptional points'' [Phys.Rev.Letters 86, 787 (2001)]
}

\author{I.~Rotter}
\address{
Max-Planck-Institut f\"ur Physik komplexer Systeme,
D-01187 Dresden, Germany}

\date{\today}

\maketitle

\vspace*{.5cm}

 ]  

Dembowski et al. \cite{richter} report on a microwave cavity experiment where
exceptional points (EP) are encircled. 
The authors claim that the EP is a
fourth order branch point for the wave functions because the two wave functions
are not just interchanged as their corresponding energies during a complete
loop in the parameter ($\lambda$) plane, but one of them undergoes a phase 
change. Therefore, the EP can, according to the authors, be clearly 
distinguished from other topological
singularities such as diabolic points (DP). The last ones are encircled some
years ago in a similar experiment \cite{dubb}. In the following, it will be
shown that the two experiments \cite{richter} and \cite{dubb} 
prove the same topological singularity in
contrast to the statement of \cite{richter}.

The encircled EP  \cite{richter,hemuro} is nothing else than a 
second order branch point in the complex plane (BPCP) 
with the following properties \cite{ro01}:\\ 
(i) the  complex energies  of the two crossing states are equal, 
$E_1 - \frac{i}{2} \Gamma_1 =  E_2 - \frac{i}{2} \Gamma_2 \; ,
$ \\
(ii) their wave functions are linearly dependent, 
\begin{eqnarray}
\psi_1 = \; \pm \; i \; \psi_2 \; , 
\label{2}
\end{eqnarray}
and bi-orthogonal with 
$\langle \psi_i| \psi_i \rangle  \to \infty$,
$|\langle \psi_i| \psi_{i\ne j} \rangle | \to  \infty $
 and $ \langle \psi_i^*| \psi_j \rangle = \delta_{i,j} $.
Although the number of BPCP is of measure zero, their influence on the
level dynamics is quite large. It can be traced to, e.g., 
the well-known avoided  crossing of discrete or  resonance states 
\cite{ro01}. The BPCP and DP are therefore related to one another.

For illustration, let us consider the complex two-by-two Hamiltonian matrix 
 \begin{eqnarray}
{\cal H} =
 \left(
\begin{array}{cc}
 e_1(\lambda) - \frac{i}{2}\gamma_1 & \omega \\
\omega  &   e_2(\lambda) -  \frac{i}{2} \gamma_2
\end{array}
\right) 
\label{4}
\end{eqnarray}
where $e_k$ and  $\gamma_k$ ($k=1,2$) are the unperturbed energies  
and widths, respectively, of the two states. The $e_k$ are assumed to 
depend on the parameter $\lambda$  in such a manner that  
the two states may cross in energy at $\lambda^{\rm cr}$ when $\omega = 0$.
The two states interact only via $\omega $ which is assumed 
in the following  to be 
independent of the parameter $\lambda$ (as the $ \gamma_k$).
The eigenvalues of  ${\cal H}$     are 
$ E_\pm - \frac{i}{2}  \Gamma_\pm  = 
(e_1 + e_2) - \frac{i}{2} \; (\gamma_1   +  \gamma_2 )
\pm \frac{1}{2} \;
\sqrt{F} $ 
with
\begin{eqnarray}
 F=
\big(( e_1 - e_2) - \frac{i}{2} \; (\gamma_1  -  \gamma_2 )\big)^2 
+ 4 \omega^2 \; .
\label{5}
\end{eqnarray}
When $ F(\lambda,\omega) = 0$ at $\lambda=\lambda^{\rm cr}$ 
(and $\omega=\omega^{\rm cr}$),
the $S$ matrix has a double pole which is a second order  BPCP.

According to  (\ref{5}), $ F = F_R + i \, F_I$ is generally 
a complex number. For real
$\omega$,   $\; e_1 = e_2$ at $\lambda=\lambda^{\rm cr}$  and 
we have to differentiate between three cases
\begin{eqnarray}
  F_R(\lambda,\omega) > 0 & \quad \to \quad  & \sqrt{F_R} = {\rm real} 
\label{eq:fri11}\\
  F_R(\lambda,\omega) = 0 & \to  &\sqrt{F_R} = 0 
\label{eq:fri12}\\
  F_R(\lambda,\omega) < 0 &  \to  &\sqrt{F_R} = {\rm imag} \; .
\label{eq:fri13}
\end{eqnarray}
The first case gives the  avoided level
crossing in energy with an exchange of the two wave functions 
at $\lambda^{\rm cr}$. The second case 
corresponds to the double pole of the $S$
matrix. In the third case, the two levels cross freely in energy
and the two states
are {\it not} exchanged at the critical value  
$\lambda^{\rm cr}$ \cite{ro01}.
In \cite{brentano}, the
two cases $F_R > 0$ and  $F_R < 0$ are studied experimentally in
a microwave cavity and called {\it
overcritical} and
{\it subcritical} coupling, respectively.

The DP is surrounded in the experiment \cite{dubb} 
in the regime of overcritical coupling along the whole way of 
encircling and  $\lambda^{\rm cr}$ is passed twice in opposite directions,
(i) $\psi_1 \to - \, i \, \psi_2, \;
\psi_2 \to + \, i \, \psi_1 $ and (ii)
$ -i \, \psi_2 \to - \psi_1, \; i \, \psi_1 \to -  \psi_2 $.
The  phase change occuring after surrounding the DP  
corresponds to the geometric phase discussed by Berry \cite{berry}.

The way of encircling  the BPCP itself passes from
a region with  overcritical coupling at $\lambda^{\rm cr}$ to another one with 
subcritical coupling at $\lambda^{\rm cr}$. 
Thus, a first full surrounding gives
$\psi_1 \to - \, i \, \psi_2, \;  \psi_2 \to + \, i \, \psi_1 $
and a second one (in the same direction)
$- \, i \, \psi_2 \to  + \psi_1, \; i \, \psi_1 \to + \psi_2 $.
That means, surrounding the BPCP twice restores   the wave functions
$\psi_i$ including their phases. The BPCP is a second order branch point  in
correspondence with the eigenvalue equation \cite{ro01}.
Encircling the BPCP in the opposite direction
gives $\psi_1 \to + \, i \, \psi_2  , \;  
\psi_2 \to - \, i \, \psi_1 $
in agreement with the experimental data of \cite{richter}.

Summarizing, 
the phase changes observed by encircling a DP and a BPCP (called EP in
\cite{richter,hemuro})  are caused by the {\it same} topological 
singularity.

\end{document}